\documentclass[journal=nalefd,manuscript=article]{achemso}

\usepackage[version=3]{mhchem} 
\usepackage[T1]{fontenc}       
\usepackage{graphicx,color, ulem}

\author{S.~d'Hollosy}
\altaffiliation{These authors contributed equally to this work.}

\author{M.~Jung}
\altaffiliation{These authors contributed equally to this work.}

\author{A.~Baumgartner}
\email{andreas.baumgartner@unibas.ch}
\affiliation[Institute of Physics, University of Basel, Klingelbergstrasse 82, 4056 Basel, Switzerland]{University of Basel}

\author{V.A.~Guzenko}
\affiliation[Laboratory for Micro- and Nanotechnology, Paul Scherrer Institute, CH-5232 Villigen PSI, Switzerland]{PSI Villigen}

\author{M.H.~Madsen}
\altaffiliation{present address: Danish Fundamental Metrology, DK-2800 Kgs. Lyngby, Denmark}

\author{J.~Nyg{\aa}rd}
\affiliation[Center for Quantum Devices and Nano-Science Center, Niels Bohr Institute, University of Copenhagen, Universitetsparken 5, DK-2100
Copenhagen, Denmark]{University of Copenhagen} 

\author{C.~Sch{\"o}nenberger}
\affiliation[Institute of Physics, University of Basel, Klingelbergstrasse 82, 4056 Basel, Switzerland]{University of Basel}

\title{Giga-Hertz quantized charge pumping in bottom gate defined InAs nanowire quantum dots}

\begin{document}

\begin{abstract}
Semiconducting nanowires (NWs) are a versatile, highly tunable material platform at the heart of many new developments in nanoscale and quantum physics. Here, we demonstrate charge pumping, i.e., the controlled transport of individual electrons through an InAs NW quantum dot (QD) device at frequencies up to $1.3\,$GHz. The QD is induced electrostatically in the NW by a series of local bottom gates in a state of the art device geometry. A periodic modulation of a single gate is enough to obtain a dc current proportional to the frequency of the modulation. The dc bias, the modulation amplitude and the gate voltages on the local gates can be used to control the number of charges conveyed per cycle. Charge pumping in InAs NWs is relevant not only in metrology as a current standard, but also opens up the opportunity to investigate a variety of exotic states of matter, e.g. Majorana modes, by single electron spectroscopy and correlation experiments.

%

\end{abstract}

Epitaxially grown semiconducting nanowires (NWs) are versatile material platforms with myriads of fundamental and practical applications.\cite{Lieber_Wang_MRS_Bull32_2007} InSb and InAs NWs are at the heart of several conceptually new experiments, e.g. in the search for Majorana bound states,\cite{Mourik_Kouwenhoven_Science_2012, Das_Heiblum_NatPhys_2012} or in a Cooper pair splitter, a potential source of entangled electron pairs.\cite{Hofstetter2009, Hofstetter_Baumgartner_PRL107_2011, Das_Heiblum_NatureComm_2012} Many NWs are optically active\cite{Li_Lieber_MaterialsToday9_2006} and can be incorporated in highly tunable device geometries.\cite{Mourik_Kouwenhoven_Science_2012, Fulup_dHollosy_Baumgartner_PRB90_2014} The spin coherence times and large g-factors are ideal to form and manipulate quantum bits.\cite{Nadj-Perge_Frolov_Bakkers_Kouwenhoven_Nature468_2010, Petersson_Jung_Petta_Nature490_2012}
New measurement paradigms like initialization and coherent control of quantum bits, dynamic quantum control, or single electron spectroscopy and correlation measurements require the fast manipulation of individual electrons on the scale of several hundred MHz. The controlled transport of single electrons can be achieved in a process called 'quantized charge pumping' (CP), in which a periodic modulation of one or several external parameters leads to a dc current of one electron per cycle.

For many years the driving force for the development of CP was the creation of a new current standard in metrology.\cite{Pekola_RevModPhys_2013} Recent theoretical proposals suggest that CP can be used well beyond a current standard, for example to investigate and characterize exotic electron states like fractional quantum Hall states.\cite{Das_Shpitalnik_EPL83_2008} Specifically, CP in NW devices was proposed to identify Majorana bound states,\cite{Gibertini_PRB_2013} or fractional fermions.\cite{Saha_Loss_arXiv_2014} Very recent experiments in two-dimensional electron gases (2DEGs) have already demonstrated an on-demand source for electron pairs with potentially non-classical correlations.\cite{Ubbelohde_Haug_NatureNanotech_2014}

CP driven by electrical gates at frequencies of up to a few MHz was demonstrated in channels of a 2DEG,\cite{Kouwenhoven_PRL67_1991} metallic single electron transistors,\cite{Pothier_Urbina_Devoret_EPL17_1992} in InAs NWs\cite{Fuhrer_Samuelson_APL91_2007} and in carbon nanotubes.\cite{Chorley_Buitelaar_APL100_2012} Higher gating frequencies became accessible recently in 2DEGs,\cite{Blumenthal_Kaestner_Ritchie_NaturePhys_2007, Giblin_NatureComm_2012} Si-based field-effect transistors\cite{Fujiwara_APL92_2008, Yamahata_APL106_2015} and graphene.\cite{Connolly_NatureNano_2013} An essential step for applications and more involved experiments was the demonstration of CP by modulating only a single external parameter.\cite{Fujiwara_APL92_2008, Kaestner_Blumenthal_PRB77_2008}

Here we present single-parameter quantized CP in InAs NW QDs defined and controlled by local bottom gates and a global backgate with a device geometry and material choices similar to the most recent advanced low-frequency NW experiments. We find current steps for frequencies up to $1.3\,$GHz, which corresponds to a dc current of $208.4\,$pA. We investigate the dependence of CP on the gate voltages, bias and applied rf-power and discuss non-equilibrium effects at high frequencies. Our results demonstrate the feasibility of using CP as a tool to investigate novel phenomena in semiconducting NWs, such as Majorana bound states or Cooper pair splitting. In addition, NWs are promising for studying non-equilibrium quantum phenomena and for improved current standards because the low effective electron mass results in a large energy level spacing.
 
The mechanism of single-parameter non-adiabatic CP is shown schematically in Fig.~1a. Local electrical gates define two tunnel barriers (left and right, L and R), forming a QD with discrete electron states separated by the Coulomb charging and level spacing energies. A central gate (M) can be used to fine-tune the QD energies. A periodic voltage modulation on gate L leads to a modulation of the left potential barrier. In the loading phase (i) of a CP cycle, the left barrier is low and an empty QD state is filled by an electron when below the Fermi energy, $E_{\rm F}$, of the leads (depicted for zero bias). In the subsequent decoupling phase (ii), the left barrier is increased, which prohibits the electron from tunneling back into lead L and lifts the filled state above $E_{\rm F}$ due to the capacitive coupling to the QD. In the unloading phase (iii) the QD state reaches above the right barrier and the electron leaves into the empty states of the right contact. In the second half of the modulation the left barrier and the now empty QD state are lowered again while the right barrier prohibits the QD from being charged from the right. At any time of the cycle the static dc conductance through the structure is negligible. At most times the electrons in the QD are not in thermal equilibrium with the contacts ('non-adiabatic') and at frequencies higher than the energy relaxation rate also the QD level occupation can be non-thermal ('hot electrons'). If each cycle transfers on average one electron per state from the left to the right, the resulting average dc current is
\begin{equation}
I_{\rm dc}=nef,
\label{eq1}
\end{equation}
with $n$ the number of electron states accessible for CP, $e$ the elementary charge and $f$ the frequency of the modulation. As demonstrated below, $n$ can be controlled by various parameters, such as gate voltages, dc bias and modulation amplitudes, which leads to steps between well-defined current levels as a function of the respective parameter.

\begin{figure}[t]
	\centering
		\includegraphics[width=0.5\textwidth]{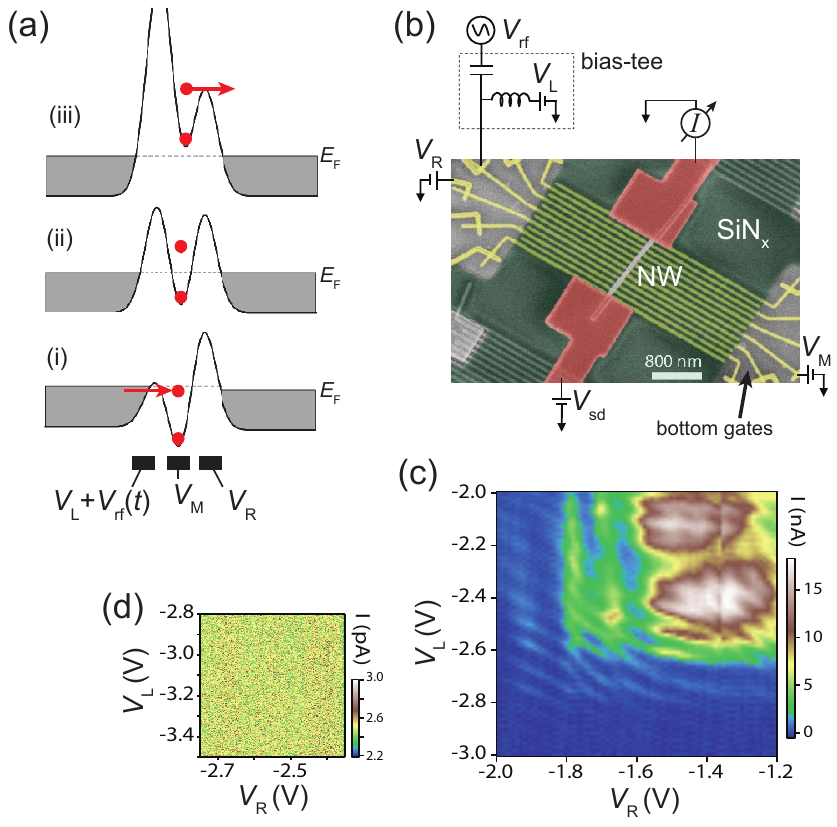}
	\caption{(Color online) (a) Schematic of the potential landscape during a charge pumping cycle. Three bottom gates are used to form a QD with the left voltage varied in time. Depicted are three phases of the modulation cycle: the (i) loading, (ii) decoupling and (iii) unloading phase. (b) False color scanning electron microscopy image of a CP device. (c) Direct current through the QD as a function of the barrier gate voltages $V_{\rm L}$ and $V_{\rm R}$. (d) The dc current is zero in the gate voltage range of the charge pumping experiments.}
\end{figure}


A scanning electron microscopy (SEM) image of a NW charge pump is shown in Fig.~1b. It consists of an array with twelve local gates fabricated by electron beam lithography on a highly doped silicon substrate that serves as a global backgate, insulated by $\sim400\,$nm SiO$_2$. The $\sim50\,$nm wide local gates consist of $4\,$nm Ti and $18\,$nm Pt with an edge to edge separation of $\sim60\,$nm. These gates are overgrown by $\sim25\,$nm SiN$_x$ for electrical insulation using plasma-enhanced chemical vapor deposition. The SiN$_x$ was removed at the ends of the gates by a reactive ion etch with CHF$_3$/O$_2$\cite{Wong1992} for electrical contact. In the next step we deposit a single InAs NW ($\sim 70\,$nm diameter) perpendicular to the gate axes using micromanipulators. The NWs were grown by solid-source molecular beam epitaxy,\cite{Madsen_Nygard_JCrystGrowth_2013} implementing a two-step growth process to suppress stacking faults.\cite{Shtrikman_Heiblum_Nanolett_2009} Two Ti/Pd $(5/100\,$nm) contacts were fabricated after an Ar plasma etch to remove the native oxide at the ends of the NW.

Figure~1b also shows the measurement setup. A QD is formed using the central three bottom gates, labeled left (L), middle (M) and right (R) with the corresponding gate voltages $V_{\rm L}$, $V_{\rm M}$ and $V_{\rm R}$. The remaining gates are held constant at $1\,$V. A standard frequency generator supplies a sinusoidal radio-frequency (rf) superimposed to the dc voltage on gate L using a commercial bias-tee at room temperature. This signal is fed into an rf line connected to the left gate. All other lines are standard twisted pair copper cables. The sample is mounted on a printed circuit board and resides in a vacuum chamber immersed in a bath of liquid helium at $4.2\,$K. A dc bias is applied on the right NW contact, while a standard room-temperature current-voltage (IV) converter is used to read out the dc current through the left NW end to ground.

The dc current through the device, $I_{\rm dc}$, without an rf-excitation is plotted in Fig.~1c as a function of the barrier gate voltages $V_{\rm L}$ and $V_{\rm R}$, with a bias of $V_{\rm sd}=1\,$mV and the middle gate set to $V_{\rm M}=-200\,$mV. Coulomb blockade resonances can be discriminated down to $V_{\rm L}\approx-3\,$V and $V_{\rm R}\approx-2\,$V, where the transport characteristics are dominated by a single QD (see supporting information, Fig.~S1). At higher voltages the system exhibits a larger conductance and a more complex pattern. From standard transport spectroscopy in the single dot regime we obtain an addition energy of $E_{\rm add}\approx 7\,$meV, with a level spacing of $\sim1\,$meV. All CP experiments presented below were obtained at strongly negative barrier gate voltages, where the dc current without CP, shown in Fig.~1d, is too small to be detected in our setup, i.e. $I_{\rm dc}<10\,$fA. The IV converter has a current offset of $2-4\,$pA not subtracted from the data.

\begin{figure}[t]
	\centering
		\includegraphics[width=0.5\textwidth]{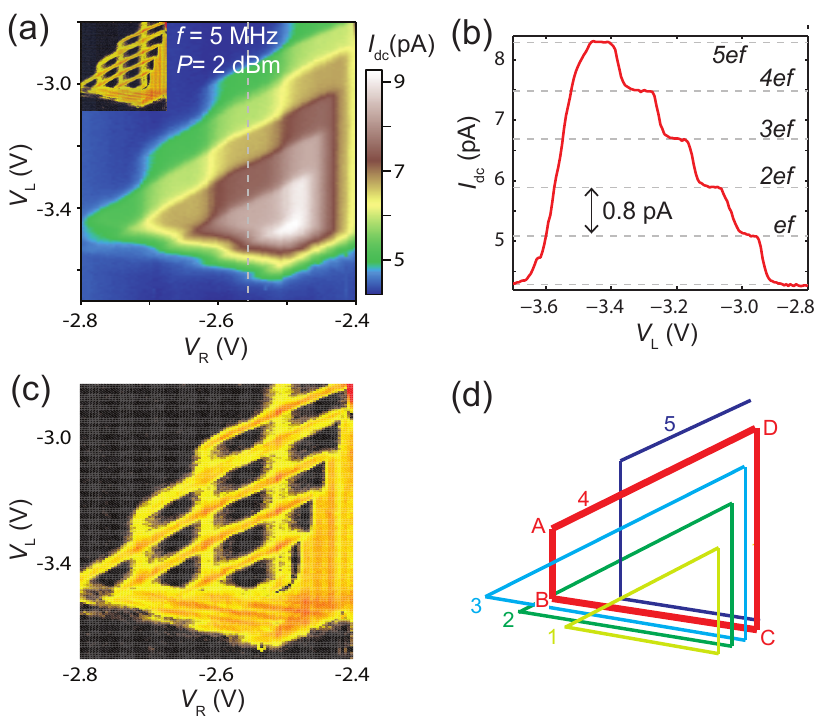}
	\caption{(Color online) (a) dc current, $I_{\rm dc}$, as a function of the left and right barrier gate voltages at $f=5\,$MHz. (b) dc current steps in the cross section indicated in Fig.~a. The expected current step values of $nef$ are indicated by horizontal dashed lines. (c) The absolute value of the current gradient with respect to the gate voltages, highlighting the current steps (also shown as an inset in a). (d) Schematic of transition lines between current plateaus.}
\end{figure}

Figure~2a shows the dc current through the NW as a function of the left and right barrier gate voltages with a sinusoidal modulation at $f=5\,$MHz and the power $P=2\,$dBm on $V_{\rm L}$. In all experiments the dc bias is set to $V_{\rm sd}=1\,$mV. The resulting current shows plateaus of constant current bounded by sharp transitions to the next higher values. 
This step structure can be seen clearly in Fig.~2b where a cross section of the data at constant $V_{\rm R}$ (dashed vertical line in Fig.~2a) is plotted as a function of $V_{\rm L}$. Five quantized current plateaus occur at $I=nef$, indicated by horizontal dashed lines. The step size is $\Delta I=0.80\,$pA at this pumping frequency, as expected from Eq.~\ref{eq1}. The accuracy of the step size is roughly $\delta I / \Delta I\approx 10^{-3}$ with $\delta I$ the mean square deviation from the theoretical value and $\Delta I$ the step size. In our setup this accuracy is limited mainly by the noise in the current measurement, which can be improved significantly by an optimized current comparator setup.

The transitions between current plateaus are more clearly visible in the current gradient with respect to the two gate voltages, as shown in Fig.~2c and in the inset of Fig.~2a. The position of the transitions is schematically shown in Fig.~2d with the onset and suppression of CP for different QD states ($1-5$).
For orientation, we briefly discuss the CP gate dependence, considering state 4 in Fig.~2d. A simple electrostatic model is described in the supporting information. CP is limited at higher $V_{\rm L}$ (line AD in Fig.~2d) when a filled QD state does not reach beyond the right barrier in the unloading process. The position of the line depends on the QD level energy, but the positive slope only on the parameters that couple the gate voltages to the level energy and the barrier height (see Eq.~S3 in the supporting information). Since the lever arms to the QD can be inferred from dc experiments, we can deduce the coupling parameter of the right gate to the height of the right barrier, which is about twice the lever arm to the QD in our experiments.
The separation of two consecutive AD lines, i.e. the extent of the quantized current plateaus in Fig.~2b, is determined solely by the addition energy of the QD and the lever arm of the left gate, see Eq.~S4 in the supporting information. This is one instance at which the small effective mass of InAs, and thus the large confinement energies, might be advantageous for metrological applications. With the QD lever arms we can directly estimate the addition energies of the {\it dynamic} QD at the minimum of the rf voltage.

When $V_{\rm L}$ becomes too negative, the QD energy levels are decoupled from the left lead before they reach below $E_{\rm F}$ in the loading phase. However, electrons tunneling into energetically higher states and subsequent energy relaxation can still lead to the occupation of the lower states. In this case the loading phase is limited for all states almost at the same gate voltages, at which the highest lying confined state that still reaches below $E_{\rm F}$ becomes occupied in the loading phase. The alignment of this highest state with $E_{\rm F}$ thus defines the limiting line BC in Fig.~2d  and leads to a negative slope identical to the gate-dependence of the QD resonances in the dc experiments (see discussion leading to Eq.~S5 in the supporting information for more details).

The opening of the triangle given by the lines AD and BC at a given $V_{\rm R}$ increases linearly with the rf voltage amplitude and $V_{\rm R}$ (Eq.~S6 in the supporting information), as in the experiment. It is this opening that determines how many current plateaus can be observed.

For the higher lying states (4 and 5 in Fig.~2d), CP is also limited at the almost vertical lines AB. Here, the electron tunnels back to the left in the decoupling phase, which suppresses CP for this state. Back-tunneling becomes suppressed at the specific time in the rf cycle when the barrier pierces $E_{\rm F}$. The potential landscape at this point in time is independent of $V_{\rm L}$ and CP is possible if the state is below $E_{\rm F}$ at this time, which only depends on $V_{\rm R}$. The QD states at lower energies (states 1-3 in Fig.~2d) are not confinement-limited because the barrier is already established when the states cross $E_{\rm F}$. In Eq.~S7 of the supporting information the spacing between consecutive AB lines is determined by the QD addition energy and the lever arm of the right gate. Assuming lever arms that do not change during the CP cycle (see Fig.~S1b in supporting information), we find that the addition energy of the QD in the unloading phase (AD) is in average almost a factor of 2 larger than in the confinement phase (AB) of the rf cycle, $E_{\rm add}^{\rm AD}/E_{\rm add}^{\rm AB}\approx 1.9$, see Eq.~S8. This result is quite intuitive because of the deeper and stronger confinement expected in the unloading phase, but contradicts many theoretical assumptions and estimates used in the literature.

CP is limited at the vertical lines CD when the right barrier is too low and the QD is not emptied reliably to the right because of back-tunneling from the right contact. In contrast to other experiments, in our measurements the CP current tends to zero and does not increase, as one might expect when both barriers are low.

\begin{figure}[t]
	\centering
		\includegraphics[width=0.5\textwidth]{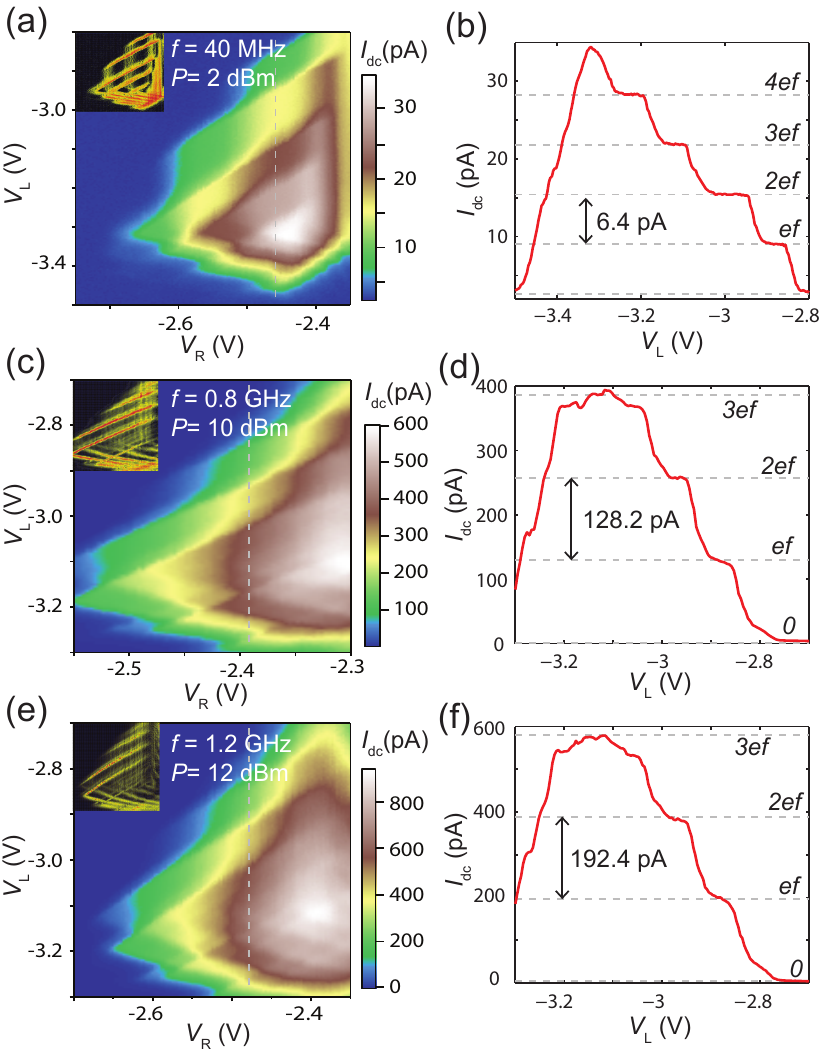}
	\caption{(Color online) Left column: dc current, $I_{\rm dc}$, as a function of the left and right barrier gate voltages for a series of rf frequencies $f$. The insets show the absolute value of the gradient in the dc current with respect to the gate voltages, which highlights the current steps. Right column: dc current steps in cross sections at a constant $V_{\rm R}$, indicated by a dashed line in the respective figure to the left. The expected current steps of $ef$ are indicated by horizontal dashed lines.}
\end{figure}

We now turn to the question of how the pattern in Fig.~2 evolves with increasing frequency.
Figure~3a shows the CP current as a function of $V_{\rm L}$ and $V_{\rm R}$ at $f=40\,$MHz, with a cross section at $V_{\rm R}=-2.46\,$V plotted in Fig.~3b. The applied rf power is $P=2\,$dBm. Also here the current steps due to CP are well defined and exhibit a similar gate dependence as at $5\,$MHz. The current step size is $\Delta I=6.4\,$pA, in agreement with ideal CP, with a relative accuracy of the current steps similar to lower frequencies. The main difference to $5\,$MHz in the gate dependence is that the vertical transitions AB appear broadened and seem to consist of several parallel features (see inset). The current steps in cross sections parallel to the lines AD (not shown) appear washed out, while the vertical cross sections still exhibit clearly defined plateaus.

At $f=800\,$MHz the dc current shown in Fig.~3c exhibits plateaus with $\Delta I=128.2\,$pA steps and well-defined transitions along the AB lines, but strongly washed out along AD. The applied rf power is $P=10\,$dBm, but damping due to stray capacitances leads to a reduced rf power on the left gate and a reduced number of visible plateaus compared to the experiments at lower frequencies. In the cross section in Fig.~3d one finds that not all plateaus at $I=ef$ are flat, but can exhibit a slope as a function of $V_{\rm L}$, and that the transition regions between plateaus exhibit kinks, most prominently between $n=0$ and $n=1$. The vertical transition lines AB in the gradient image are separated into several weaker lines of almost equal spacing, whereas the lines AD and BC seem as strong as at lower frequencies. Most prominently, the spacing of the BC lines is almost doubled and reaches almost the spacing of the AD lines for some transitions.

The picture is similar for $f=1.2\,$GHz shown in Figs.~3e and f, with an rf power of $P=12\,$dBm. Here, the rf-power losses are considerable, which can be traced back mainly to the capacitance of the close-by metallic bottom-gates and losses in the setup. The current steps in Fig.~3f are $\Delta I=192.4\,$pA, but the plateaus now show a significant slope. Cross sections along the AD lines (not shown) only exhibit very vague modulations on top of a steady increase in the current. The spacing between the weak AB lines remains essentially unchanged compared to $800\,$MHz.

The high frequency CP experiments can be understood qualitatively in our simple model. Similar additional vertical AB lines for high frequencies were reported recently\cite{Kataoka_Ritchie_PRL106_2011} and attributed to the non-equilibrium occupation of excited states when the rf frequency is larger than the relaxation rate in the QD. This leads to plateaus determined by the level spacing and not by the larger Coulomb interaction energy. Also the slope of the current plateaus was attributed theoretically to such non-thermal electron distributions.\cite{Flensberg_PRB60_1999} Our experiments show that this distribution mainly affects the separation phase in which electrons in excited states can tunnel back into the left contact more easily due to the lower barrier when crossing the Fermi energy. This leads to the observed smearing of the lines along AB and also accounts qualitatively for the sloped plateaus. We expect that this smearing would be considerably stronger for materials with a larger effective mass.
Another non-equilibrium effect is the increasing separation of the BC lines with increasing frequency. If the time window of the loading process becomes shorter than the electron energy relaxation time, the states can only be charged by direct tunneling when aligned with $E_{\rm F}$ at more positive $V_{\rm L}$, in contrast to low frequency CP, where energy relaxation allows the loading of the ground state at lower $V_{\rm L}$ via excited states. This results in a spacing of the BC lines determined by the addition energy, similar to the AD lines, in rough agreement with the experiments at higher frequencies.

\begin{figure}[t]
	\centering
		\includegraphics[width=0.5\textwidth]{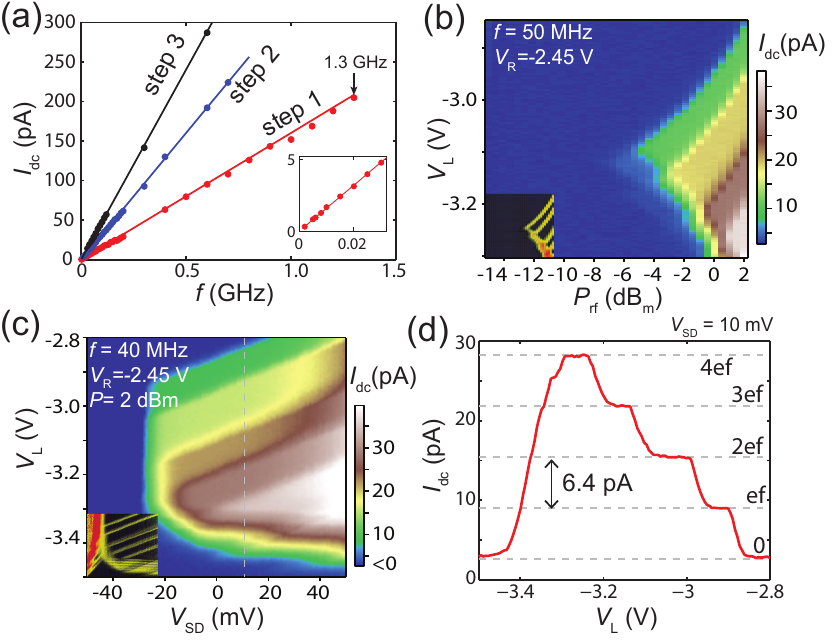}
	\caption{(Color online) (a) dc current plateau values as a function of the applied rf frequency. An offset of $4\,$pA was subtracted for all points. The straight lines are given by $I_{\rm dc}=nef$ without adjustable parameters. The inset shows the denser low-frequency points. (b) $I_{\rm dc}$ as a function of the applied rf-power $P_{\rm rf}$ and the left gate voltage, $V_{\rm L}$. (c) $I_{\rm dc}$ as a function of the bias, $V_{\rm sd}$, and $V_{\rm L}$. The colorscale is limited to positive values. For $V_{\rm sd}<-23\,$mV the current is negative and increases continuously in amplitude. The insets of the colorscale images are the absolute values of the gradient of $I_{\rm dc}$. (d) Cross section of (c) at $V_{\rm sd}=10\,$mV}.
\end{figure}

We now investigate the CP current as a function of other external parameters. The frequency dependence of the dc current plateaus is plotted in Fig.~4a. The symbols are the experimental values, while the straight lines are given by $I_{\rm dc}=nef$, where $n$ is the number of electrons conveyed per rf cycle, without any adjustable parameters. We clearly find the expected linear frequency dependence of the CP current up to the highest investigated value of $1.3\,$GHz. The systematic deviations at higher frequencies are due to the slope in the plateaus, for which we chose the central value of the plateau, which systematically underestimates the plateau value because the non-equilibrium electrons in the excited states are more probable to tunnel back into the left contact, which reduces the CP current.\cite{Kataoka_Ritchie_PRL106_2011}

The dependence of the CP current on the rf power $P$ is investigated in Fig.~4b for $V_{\rm L}$-sweeps at $f=50\,$MHz and constant $V_{\rm R}=-2.45\,$V. We cannot detect CP for powers below $-8\,$dBm. With increasing power CP is limited to an increasing interval in $V_{\rm L}$. The more positive $V_{\rm L}$-limit is determined by the unloading process (line AD in Fig.~2d), and the more negative by the loading-limited BC line. The evolution of these lines with power is very well described by Eqs.~S5 of our simple model in the supporting information, using the conversion of power to the applied voltage $V_{\rm rf} [{\rm mV}]=0.22\times 10^{P[{\rm dBm}]/20}$, and a frequency-dependent prefactor. The latter is about $1$ at the lower frequencies shown here and decreases for higher frequencies. These experiments can in principle be used directly to calibrate the power applied to the gates.

In Fig.~4c we plot the CP current as a function of $V_{\rm L}$ and the bias, $V_{\rm sd}$, applied to the right contact, for $f=40\,$MHz. A cross section at  $V_{\rm sd}=10\,$mV is shown in Fig.~3d. The colorscale is limited to positive values because for $V_{\rm sd}<-23\,$mV the current evolves continuously, i.e., without steps, to very strongly negative values. For $V_{\rm sd}>-23\,$mV we find a positive current and very sharp and well defined current plateaus, demonstrating that electrons can also be pumped against a significant bias in our NW pumps. The break-down of CP and the strong negative current for a strongly negative bias is due to the bias distorting the left tunnel barrier and the left Fermi energy reaching above the left potential barrier at least at the maximum voltage of the rf cycle. Since the current steps cease roughly at the same bias, the effect by the capacitive coupling from the left contact to the QD can be neglected. However, this source capacitance is responsible for the step wise change in $I_{\rm dc}$ with the bias at more positive $V_{\rm L}$.


In summary, we report the fabrication and detailed characterization of an InAs NW based single-parameter charge pumping device up to GHz frequencies. We discuss in detail the dependencies on rf power, bias and gate voltages, which is possible due to the mutiple-gate design of the device. On the one hand, this proof of concept opens up the important material category of NWs, with high quality crystals and low effective mass, to be used in current standard experiments. Also the straight forward scalability in NW arrays to reduce the error rate in parallel charge pumping devices is worth considering. In addition, since NWs are optically active, one can also envisage to convert the CP current into a standardized photon flux, thus linking the light intensity to a frequency and current standard. On the other hand, our experiments demonstrate the feasibility to use CP as a novel spectroscopy tool to investigate correlated electron systems and exotic transport processes in NWs, e.g., in a Cooper pair splitter or Majorana zero modes, where CP might reduce background fluctuations like shot noise to improve the visibility of cross correlation signals.

\begin{suppinfo}
\begin{itemize}
	\item Quantum dot characteristics in the single-dot regime
	\item Electrostatic model of quantized charge pumping
\end{itemize}
\end{suppinfo}

\begin{acknowledgement}
We thank G. Fabian, M. Seo, M.-H. Bae and N. Kim for helpful discussions. This work was supported by the EU ERC project QUEST, the EU FP7 project SE2ND, the Swiss National Science Foundation (SNF), including the project NCCR QSIT, the Danish Research Councils and the Innovation Fund (DSF).
\end{acknowledgement}

\section{Supplementary material}

\renewcommand{\thefigure}{S\arabic{figure}}
\renewcommand{\theequation}{S\arabic{equation}}

\section{Quantum dot characteristics}

\begin{figure}[b!]
	\centering
		\includegraphics[width=0.5\textwidth]{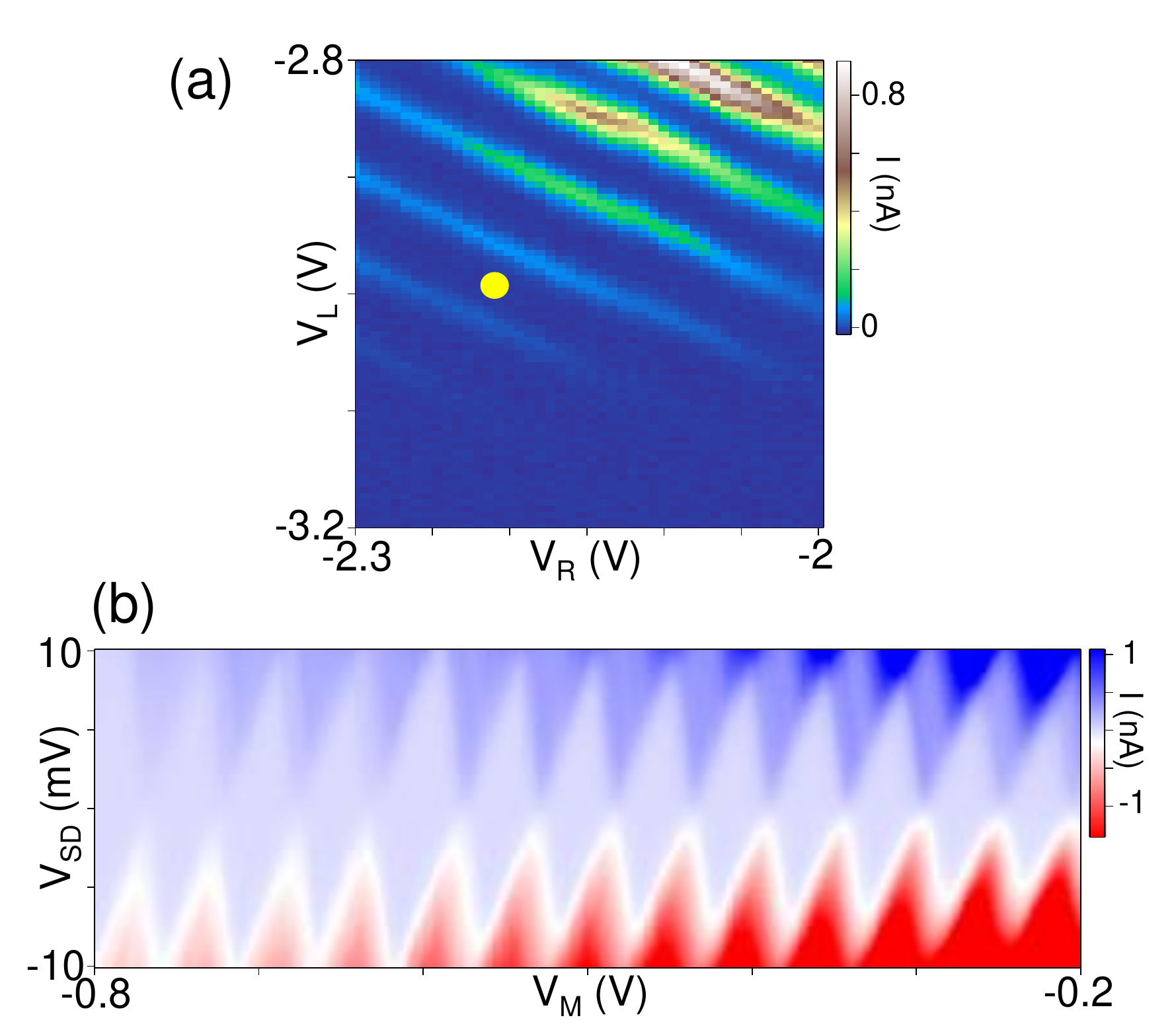}
	\caption{(Color online) (a) dc current as a function of the right and left barrier gate voltages, $V_{\rm R}$ and $V_{\rm L}$, respectively, at zero bias. The (yellow) point indicates the gate voltages at which the measurements in (b) were recorded. (b) dc current as a function of the middle gate voltage, $V_{\rm M}$ and the bias $V_{\rm SD}$.}
\end{figure}

In Fig.~S1a the current though the quantum dot (QD) is plotted as a function of the two barrier gate voltages $V_{\rm R}$ and $V_{\rm L}$ for the intervals where the device exhibits single-QD characteristics. The current in this regime is too small to be clearly discriminated in Fig.~1b of the main text. The Coulomb blockade resonance lines are almost straight, a first indication that the device is dominated by a single QD. Fig.~S1b shows the dc current as a function of the middle gate voltage, $V_{\rm M}$, and the bias, at barrier gate voltages corresponding to the (yellow) point in Fig.~S1a. Clear-cut periodic Coulomb diamonds and excited states with identical slopes for all diamond edges clearly support the single-QD picture. At $V_{\rm M}=-200\,$mV we find an addition energy of $\sim 7\,$meV and a lever arm of the middle gate of $\alpha_{\rm M}\approx 0.17$. We point out that the lever arm to the QD does not change between the observed diamonds. The lever arms of the neighboring gates are typically smaller by a factor of $2-3.5$. Relevant in the main text is the ratio $\alpha_{\rm L}/\alpha_{\rm R}\approx 1.67$, which can be extracted directly from Fig.~S1a. The slightly increasing addition energy at more negative $V_{\rm M}$ suggests that the QD is filled only by few electrons. 

\section{Simple electrostatic model for quantized charge pumping}

The basic structure of the gate voltage, power and bias dependence of the CP process can be understood in a simple electrostatic model. We assume that the tunnel barrier heights scale linearly with the barrier gate voltages
\begin{align}
\phi_{\rm R}&=\phi_{\rm R}^{(0)}-e\beta_{\rm R}V_{\rm R} \nonumber \\
\phi_{\rm L}&=\phi_{\rm L}^{(0)}-e\beta_{\rm L}\left[V_{\rm L}+V_{\rm rf}\sin(\omega t)\right],
\label{eq2}
\end{align}
with the constants $\beta_{j}$ describing the effect of the voltage on the barrier height, $\phi_{j}^{(0)}$ the height at zero voltage and $V_{\rm rf}$ the amplitude of the rf voltage modulation. Similarly, the chemical potential of the quantum dot with $N$ electrons depends on the two gate voltages as
\begin{equation}
\mu_{N}=\mu_{N}^{(0)}-e\alpha_{\rm L}\left[V_{\rm L}+V_{\rm rf}\sin(\omega t)\right] -e\alpha_{\rm R}V_{\rm R},
\label{eq3}
\end{equation}
with the usual lever arms $\alpha_{j}$, the chemical potential at zero voltage $\mu_{N}^{(0)}$ and the cycle frequency $\omega$. We have omitted the terms for the other gate voltages and the bias for clarity. We use the chemical potential to define the addition energy $E_{\rm add}=\mu_{N+1}^{(0)}-\mu_{N}^{(0)}$.

The onset of CP based on state $N$ with more negative $V_{\rm L}$ occurs when at the minimum rf voltage the chemical potential of the QD is maximum and lifted beyond the right barrier in the unloading process, $\mu_{N}^{\rm max}=\phi_{\rm R}$. We note that tunneling already sets in at energies slightly below the barrier maximum, but we omit such offsets for clarity. This leads to the parametrization of line AD in Fig.~2d as
\begin{equation}
V_{\rm L}^{\rm AD}=V_{\rm rf}-\frac{\phi_{\rm R}^{(0)}-\mu_{N}^{(0)}}{e\alpha_{\rm L}} + \frac{\beta_{\rm R}-\alpha_{\rm R}}{\alpha_{\rm L}}V_{\rm R}^{\rm AD}.
\label{eq4}
\end{equation}
The slope of the AD line therefore is $\frac{dV_{\rm L}^{\rm AD}}{dV_{\rm R}^{\rm AD}}=\frac{\beta_{\rm R}-\alpha_{\rm R}}{\alpha_{\rm L}}$, which is positive in the experiment and this suggests $\beta_{\rm R}>\alpha_{\rm R}$ (here by about a factor of 2).

Two AD lines originating from two consecutive states, $N$ and $N+1$, are spaced by $\Delta V_{\rm L}$ at a constant $V_{\rm R}$, which determines the plateau width in the left gate voltage, as measured in the cross section plots of Figs.~2, 3 and 4. In our simple model one finds
\begin{equation}
\Delta V_{\rm L}^{\rm AD}=V_{\rm L}^{N+1}-V_{\rm L}^{N}=\frac{\mu_{N+1}-\mu_{N}}{e\alpha_{\rm L}}=\frac{E_{\rm add}}{e\alpha_{\rm L}}.
\label{eq5}
\end{equation}

If $V_{\rm L}$ is made more negative, CP is limited by the loading process, which leads to the line BC in Fig.~2d (see main text).
If the QD levels were in thermal equilibrium with the left Fermi energy for the charging cycle, we would expect a BC line spacing determined by the addition energy and the left gate lever arm, the same way as in the unloading phase. However, if the left barrier is still large at the onset of the loading phase (this is the relevant case since the back-tunneling would null CP otherwise), the states cannot be charged by direct tunneling from the left Fermi sea. The most probable charging mechanism then becomes tunneling into the highest QD excited state that still reaches below $E_{\rm F}$ in the charging phase and subsequent relaxation to the ground state. In this picture the charging lines are much closer spaced than expected by the addition energy, intuitively separated by the level spacing.
The QD is filled as long as the highest lying excited state at $E_{\rm k}$ reaches below the Fermi energy $E_{\rm F}$, at the maximum of the rf cycle (minimum potential energy). The condition $E_{\rm k}=E_{\rm F}$ leads to the parametrization of the line BC as
\begin{equation}
V_{\rm L}^{\rm BC}=-V_{\rm rf}-\frac{E_{\rm F}-E_{\rm k}}{e\alpha_{\rm L}}-\frac{\alpha_{\rm R}}{\alpha_{\rm L}}V_{\rm R}^{\rm BC}.
\label{eq6}
\end{equation}
The slope of the BC line is thus given by $\frac{dV_{\rm L}^{\rm BC}}{dV_{\rm R}^{\rm BC}}=-\frac{\alpha_{\rm R}}{\alpha_{\rm L}}$, which is negative and coincides with the slope of the QD resonances in the dc characterization in Fig.~1c.

The opening of the triangle between the lines AD and BC at constant $V_{\rm R}$ can be obtained similarly as
\begin{align}
\Delta V_{\rm L}=&V_{\rm L}^{\rm AD}-V_{\rm L}^{\rm BC} \nonumber \\
=&2V_{\rm rf}-\frac{\phi_{\rm R}^{(0)}-E_{\rm F}}{e\alpha_{\rm L}}-\frac{E_{\rm k}^{\rm (0)}-\mu_{N}^{\rm (0)}}{e\alpha_{\rm L}} \label{eq7}\\
&+\frac{\beta_{\rm R}}{\alpha_{\rm L}}V_{\rm R},\nonumber
\end{align}
which is proportional to the rf-power and $V_{\rm R}$.

The transition lines AB occur only for the higher lying states and stem from the condition that electrons loaded into the QD need to be confined up to the unloading phase, i.e. they should not tunnel back into the left lead. In the non-adiabatic single parameter CP discussed here, the back-tunneling is suppressed if the QD state is below the $E_{\rm F}$ on the left, and if the tunnel barrier to the left is larger than the QD state energy. Since the effct of the modulation voltage is larger on the barrier height than on the QD energy levels ($\beta_{\rm L}>\alpha_{\rm L}$), these conditions can be merged into one: an electron stays confined during the separation phase if the energy level is below $E_{\rm F}$ at the time $t^{*}$ in the cycle at which the barrier height reaches $E_{\rm F}$, $\mu_{N}(t^{*})<E_{\rm F}$. Since this potential is independent of the dc value of $V_{\rm L}$, this transition is essentially independent of $V_{\rm L}$. Inserting in the above expressions, we find CP for $V_{\rm R}^{\rm AB} \geq \frac{\mu_{N}^{(0)}-E_{\rm F}}{e\alpha_{\rm R}}-\frac{\alpha_{\rm L}}{e\alpha_{\rm R}\beta_{\rm L}}(\phi_{\rm L}^{(0)}-E_{\rm F})$ and the spacing between two vertical AB lines as
\begin{equation}
\Delta V_{\rm R}^{\rm AB}=\frac{\mu_{N+1}^{(0)}-\mu_{N}^{(0)}}{e\alpha_{\rm R}}=\frac{E_{\rm add}}{e\alpha_{\rm R}}.
\label{eq8}
\end{equation}

Assuming constant lever arms during an rf cycle, the spacing of the AD and AB lines given in Eqs~S4 and S7 can be used to extract the ratio between the addition energies in different phases of the rf cycle:
\begin{equation}
\frac{E_{\rm add}^{\rm AD}}{E_{\rm add}^{\rm AB}}
=
\frac{\alpha_{\rm L}}{\alpha_{\rm R}}
\frac{\Delta V_{\rm L}^{\rm AD}}{\Delta V_{\rm R}^{\rm AB}}.
\label{eq9}
\end{equation}

\end{document}